\documentclass[aps,prb,twocolumn,nofootinbib, superscriptaddress,10pt,floatfix]{revtex4-2}
\usepackage{amsmath,amsthm,amsfonts,amssymb,amscd,bbold,mathrsfs,bm}
\usepackage{graphicx}
\usepackage{dcolumn}
\usepackage{physics}
\usepackage{hyperref}
\usepackage{xcolor}
\usepackage[columnwise]{lineno}
\begin{document}
\title{Annual-modulation fingerprint of the axion wind induced sideband triplet in quantum dot spin qubit sensors}
\author{Xiangjun Tan}
\email{xinshijietxj@gmail.com}
\affiliation{Department of Physics and Astronomy, University College London, WC1E 6BT London, United Kingdom}
\affiliation{Center on Frontiers of Computing Studies, Peking University, 100871 Beijing, China}
\author{Zhanning Wang}
\affiliation{School of Physics, University of New South Wales, Sydney 2052, Australia}
\date{\today}

\begin{abstract}
We propose a phase-coherent, narrowband magnetometer for searching couplings between axions and axion-like particles (ALPs) and electron spins, using gate-defined silicon quantum dot spin qubits.
With repeated Ramsey echo sequences and dispersive readout, the qubit precession response can be tracked with sub-Hz spectral resolution.
The accessible axion mass window is determined using a series of filtering protocols with consideration of sensing noise (including readout errors and $1/f$ noise).
We demonstrate clear indication of sidereal modulation of the signal due to Earth's rotation, while Earth's orbital motion induced annual amplitude envelope that generates sidebands at fixed frequency spacing $\pm \Omega_\oplus$ around the sidereal component, for axion mass between 1-10 $\mu$eV, our proposed method covers $g_{ae}$ from $10^{-14}$ to $10^{-10}$.
Incorporating this daily and annual modulation pattern in a likelihood analysis enhances the rejection of stationary or instrumental noise.
Our analysis indicates that spin-qubit magnetometry can achieve sensitivities approaching those suggested by astrophysical considerations, thus providing a complementary, laboratory-based probe of axion–electron interactions.
Although we focus on silicon spin qubit architectures, the approach is broadly applicable to spin-based quantum sensors.
\end{abstract}
\maketitle

%========================================
%========================================
\section{Introduction}
\label{Sec1: Introduction}

Axions and ALPs have attracted increasing attention as potential candidates for dark matter and, in some models, dark energy, as well as possible origins of several cosmological anomalies revealed in recent experiments \cite{Niemeyer2020, Jiang2021, Zitong2024,Talebian2024, Sikivie2024, Luu2025, Mariam2025, WangBao2025}.
The original axion model, initially proposed to solve the strong-CP problem and as a candidate for cold dark matter, establishes a clear connection between the axion mass and the Peccei-Quinn symmetry breaking scale, which simultaneously controls its couplings \cite{Peccei1977, Wilczek1978, Weinberg1978, Kim1979, Dine1981, Sikivie1983, Raffelt2008, Kim2010, Kim2018}.
In contrast, ALPs, which arise from various effective field theories and string theories, allow a much broader mass range with no fixed relation between mass and coupling strength \cite{John1983, Svrcek2006, Arvanitaki2010, Andreas2012, Luca2020}.
Pioneering theoretical works have revealed several important mechanisms for axion and ALP searches, such as axion–gluon interactions that dynamically eliminates the strong-CP term, photon–axion conversions in external magnetic fields, and spin precession induced by axion–fermion interactions \cite{Zhitnitskii1980, Raffelt1988, Visinelli2009, Andreas2012, Budker2014, Stadnik2014, Stadnik2015, David2017, Igor2018, Majorana2022, Pankratov2025}.
These studies have been considered to address or constrain many anomalies in astroparticle observations, such as stellar cooling excesses, anomalous transparency of the Universe to very-high-energy (VHE) $\gamma$-rays, small-scale structure tensions including core–cusp and satellite-abundance problems from black hole superradiance \cite{Angelis2007, Arvanitaki2011, Viaux2013, Meyer2014, Giannotti2016, Giannotti2017, Schive2014, Fermi2016, Brito2020}.

Thanks to constant experimental efforts, a wide range of theories about axion and ALPs have been tested on many platforms \cite{Graham2015, Irastorza2018, Luca2020, Choi2021, Yuanhong2022, Haowen2024, Berlin2025}.
Exclusion limits of many important parameters and coupling strength have been steadily improved, paving the way towards potential detections of axion and ALPs.
In the haloscope-type experiments and their variations (like the integration with squeezed state receivers or dielectric boosters), which mainly probe axion–photon interactions, the coupling strength has been constrained across broad range of axion mass as shown in Ref.\,\cite{QUAX2020, ABRACADABRA2021, QUAX2024, CAPP2024, ORGAN2025, HAYSTAC2025, ADMX2025, Braggio2025}.
Superconducting circuits and SQUID solid state devices, taking the advantage of the high precision quantum magnetometers, provide an alternative probes to axion-induced magnetic flux or electric fields, and are also sensitive to axion-nucleon and axion-electron couplings \cite{Asztalos2010, Budker2014, Berlin2020, ADMXA2020, Devlin2021}.
The helioscope route focuses on the detection of axions produced in the Sun: in a strong magnetic field, solar axions are converted into X-rays which can be directly measured by aligning a superconducting magnets with the Sun \cite{CAST2024, IAXO2025}.
If we consider the axion dark matter as a time-varying effective magnetic field, it can drive the electron or nuclear spin precession providing sensitivity to axion-nucleon and axion-electron couplings, which fits into the nuclear magnetic resonance (NMR) type schemes well \cite{CASPE2018, Kim2018, Christina2022}.
Finally, optical systems have been considered to probe axion-photon interaction, e.g., through searchs for vacuum magnetic birefringence and dichroism using high-intensity lasers \cite{PVLAS2016, PVLAS2020}.

To support the active search for axions and axion-like particles, metrological strategies are being actively developed to enable high-precision, high-fidelity detection of axion signals and associated induced electromagnetic effects \cite{Zhang2015, Luca2018, Derevianko2018, Salemi2021, Brady2022, Sushkov2023}.
Under this background, integrating axion detection with qubit platforms has attracted increasing attention.
Rapid advances in quantum computation and sensing have established techniques for initial-state preparation, precise qubit control, high-fidelity readout, and developing quantum error correction and mitigation across a variety of platforms, including superconducting circuits, Rydberg and neutral atoms, photonic systems, and semiconductor solid-state devices \cite{Burkard2023, Oakes2023, Zwerver2023, Chen2024, Braggio2025}.
Among these possible routes, semiconductor quantum dot spin qubits, where the spin degree of freedom of a confined electron encodes the quantum information, provide a natural platform for probing spin dynamics driven by axion-induced magnetic fields \cite{Loss1998, Morello2010, Pla2012, West2019}.
Furthermore, advances in device geometry designs, isotope purification, and strain engineering now enable systematic optimization of qubit operation, achieving a long relaxation and dephasing times \cite{Zwanenburg2013, Sigillito2019, Degli2024, Kam2024}.
These progress also open up possibilities for multi-qubit entanglement, including the preparation of squeezed, Bell, and Greenberger–Horne–Zeilinger (GHZ) states, which can further enhance sensitivity \cite{Tianyu2021, Hendrickx2021, Pont2024, Huet2025}.

Taking advantage of the phase-coherent spin responses of silicon CMOS spin qubits to axion-induced effective magnetic fields, we predicate several frequency modulations (FMs) sideband of the qubit spin polarization, enabled by a pulse-sequence noise filtering and spectral engineering techniques building on Ref.~\cite{Tan2025}.
These modulations scale with the coupling $g_{ae}$ and depend on the sensor orientation and noise, while the accessible axion mass window is determined by the sequence filtering, coherence times, and readout noise.
To further improve the search efficiency and background rejection, we identify another robust feature: an annual-modulation fingerprint appearing with parameter-free frequency spacing, appearing around the sidereal baseband component and originating from Earth's orbital modulation of the axion wind.
The manuscript is organized as follows: we first introduce the theoretical description of the qubit response to axion fields, then derive the triplet sideband structure and its amplitudes associated with the annual modulation.
Then, we present the corresponding amplitudes of the spin resonance peak under different situations.
We conclude with possible improvements and a brief outlook.

%========================================
%========================================
\section{Model and Methodology}
\label{sec:model}

In our model, we first consider the axion and ALPs as cold dark matter, whose coherent length is about $\lambda_{a} \approx (m_a v_a)^{-1}$ and much larger than the silicon qubit probe.
The mass of the axion is denoted by $m_a$, and the axion velocity is taken to be $v_a=10^{-3}$.
We treat the axion field as a classical oscillating pseudoscalar field described by $a(\bm{x},t) = a_0 \cos(m_a t - \bm{k}\cdot\bm{x}+\phi)$, where $\bm{k}=m_a \bm{v}_a$, and $\phi$ is an arbitrary phase angle.
The field amplitude $a_0$ can be evaluated using the local dark matter density: $\rho_{\text{DM}} = m_a^2 a_0^2 /2$, where $\rho_{\text{DM}} = 0.4$ GeV/cm${}^3$.

Next we introduce the probe, silicon spin qubits, for the detection of the axion- and ALP- induced effective magnetic field.
We consider an electron confined in a Si${}^{28}$-enriched, gate-defined metal-oxide-semiconductor (MOS) quantum dot system, subjected to a static external magnetic field $B_z = 0.5$ T \cite{Morello2010}.
The spin qubit wavefunction is mainly localized near the Si-SiO${}_2$ interface, allowing for the electric control via the metallic gate.
Recent fabrication advances have improved the dephasing time $T_2^*$ to the microsecond level, and the relaxation time $T_1$ to the second level \cite{Veldhorst2014, Shulman2014, Erika2016}.
Such CMOS-compatible architectures are highly aligned with industrial productions, and therefore can be scaled to two-dimensional qubit array easily \cite{Gonzalez2021, Eggli2025}.
Here, we adopt an effective qubit model, and put more attention to the interaction between the electron spin and axions.
Following the derivations and descriptions in Ref.~\cite{Tan2025}, the qubit can be described by $H_{\text{qubit}} = \omega_0 \sigma_z/2$, where $\sigma_z$ is the standard Pauli-Z matrix.
We emphasize that the qubit Larmor frequency $\omega_0$ depends not only on the Zeeman splitting from the static external magnetic field, but also on the energy renormalization due to the in-plane confinement, spin-orbit couplings, and orbital magnetic field terms.
According to Ref.~\cite{Graham2013, Stadnik2014}, the derivative coupling between the axion and fermions is: $\mathcal{L}_{\text{int}} = C_e\partial_\mu a \bar{\psi}_e \gamma^\mu \gamma^5 \psi_e/2f_a$, where $C_e$ is the dimensionless coupling constant, $f_a$ is the axion decay constant, $\bar{\psi}$ is the electron field.
Expanding the electron field and performing the Foldy-Wouthuysen transformation as described in Ref.~\cite{Stadnik2014}, neglecting the time component, the effective low-energy Hamiltonian will read $H_{\text{a}} = g_{ae} (\nabla a) \cdot \bm{\sigma}/2m_e$, where $\bm{\sigma} = (\sigma_x, \sigma_y, \sigma_z)$.
Furthermore, using the electron equation of motion, we can re-write the interaction Lagrangian as $\mathcal{L}=-i g_{ae} a \bar{\psi} \gamma_5 \psi$ with $g_{ae}=C_e m_e/f_a$.
In the non-relativistic limit, the interaction Hamiltonian becomes
\begin{equation}
    H_{\text{int}}=\frac{g_{ae}}{2m_e} (\nabla a) \cdot \bm{\sigma} = \frac{\gamma_e}{2} \bm{\sigma} \cdot \bm{B}_{\text{eff}} \,,
\end{equation}
where $\gamma_e$ is the gyromagnetic ratio of the electron.
The axion-induced effective magnetic field is:
\begin{equation}
    \bm{B}_{\text{eff}} = \frac{g_{ae}}{m_e\gamma_e} \nabla a \,.
\end{equation}
Substituting the axion field in, we get the magnitude of the effective magnetic field, which sets the primary detection limit and signal-to-noise (SNR) ratio in the qubit probe design:
\begin{equation}
    B_{\text{eff}} = \frac{g_{ae}}{m_e\gamma_e} v_a \sqrt{2\rho_{\text{DM}}} \,.
\end{equation}
Taking a typical coupling strength of order $g_{ae} \sim 10^{-13}$, the magnitude of the effective field is about $10^{-21}$ T.

%==========
\begin{figure}[htb]
\centering
\includegraphics[width=\columnwidth]{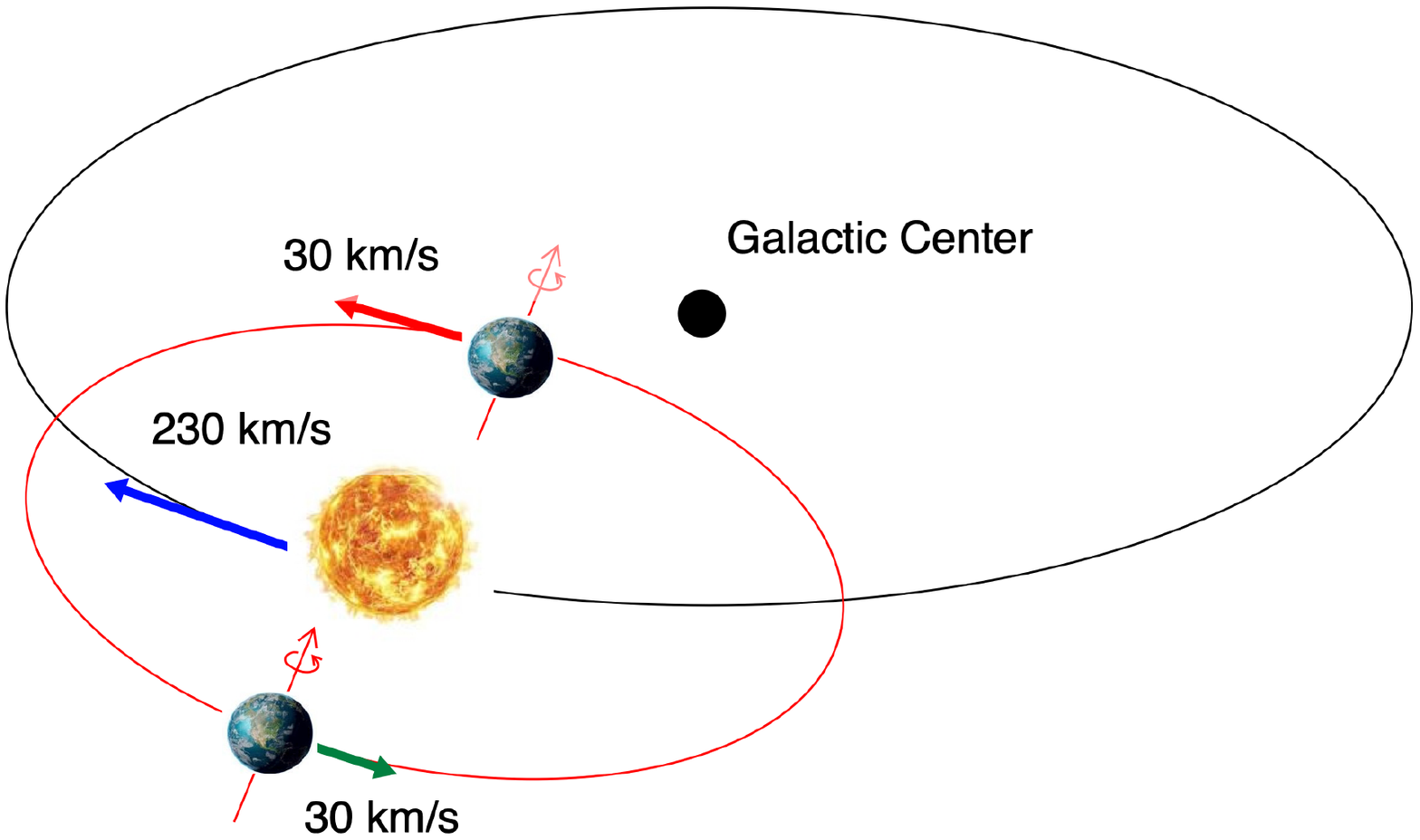}
\caption{
\textbf{Relative motion of the Solar System and Earth}.
This is a demonstration of the relative motion to study the annual modulation of the axion signal.
We consider the motion of the Solar System with respect to the Galactic rest frame as well as the orbital motion of the Earth around the Sun.
The Solar System is assumed to move with a velocity of approximately $230$ km/s relative to the Galactic center, consistent with Standard Halo Model estimates.
The Earth follows an approximately circular orbit around the Sun with velocity $30$ km/s.
The Earth’s velocity vector changes direction over the course of a year, leading to an annual modulation in the net velocity relative to the Galactic dark matter halo.
The effective axion wind observed in the laboratory frame varies periodically.
}
\label{Fig: demonstration}
\end{figure}
%==========

In the meantime, the finite coherence time of the axion field must be considered to improve the precision in estimating the SNR.
The local axion field can be treated as a narrowband random process composed of many plane waves.
The velocity dispersion therefore induces a finite spectral width broadening in axion signal spectrum.
Adopting the Standard Halo Model (SHM, as considered in Fig.~\ref{Fig: demonstration}) with a truncated Maxwell-Boltzmann speed distribution $f(v_a) \propto v_a^2 e^{-v_a^2/v_0^2} \Theta(v_{\text{esc}}-v_a)$, where $v_0$ is the most probable speed in SHM and $v_{\text{esc}}$ is the galactic escape velocity \cite{Lewin1996, Evans2019}, the mean-square speed after normalization is:
\begin{align}\label{Eq: v2-closed}
    \langle v_a^2 \rangle = v_0^2\left[\frac{3}{2} - \frac{2 z^3 e^{-z^2}}{\sqrt{\pi} N(z)}\right] \,,
\end{align}
with $N(z)=\operatorname{erf}(z) - 2 ze^{-z^2}/\sqrt{\pi}$, and $z=v_{\text{esc}}/v_0$.
The corresponding fractional linewidth of axion oscillation frequency $\nu_a = m_a/2\pi$(In natural units $m_a=\omega_a$) and axion coherence time $\tau_a$ are $\Delta\nu_a/\nu_a = \langle v^2\rangle/2$, and $\tau^{-1}_a = m_a \pi \langle v^2\rangle /2$.
For typical values $v_0=230$ km/s and $v_{\text{esc}}=544$ km/s, we find $\Delta\nu_a/\nu_a \simeq 3.9\times 10^{-7}$ and
\begin{equation}\label{Eq: tau_a}
    \tau_a \simeq 3.3~\mathrm{ms}\times\left(\frac{1~\mu\mathrm{eV}}{m_a}\right) \,.
\end{equation}
{
Coherent matched filtering yields SNR $\propto\sqrt{T_{\text{coh}}}$, with the effective coherence time defined as $T_{\text{coh}}=\min \left\{T_{\text {seg }}, \tau_a\right\}$, where $T_{\text{seg}}$ is the coherent integration segment length adjustable in experiments.
To preserve phase coherence within each matched-filter window, we restrict the segment duration to be shorter than the axion-field coherence time, i.e. $T_{\text{seg}}<\tau_a$.
Beyond this limit the stochastic phase drift of the axion field causes destructive interference in coherent sums.
In practice, the total observation is divided into multiple coherent segments of length $T_{\text{seg}}(\nu)=\min \left\{\varepsilon\tau_a(\nu), T_{\text{cap}}\right\}$.
where $\varepsilon<1$ is a safety factor and $T_{\text{cap}}$ denotes the maximum coherent capture time per segment.
The segments are then combined non-coherently in power, so the overall sensitivity continues to scale with the full observing time as $\propto \sqrt{T_{\text{tot}}}$.
This choice also keeps the annually/sidereally modulated signal approximately stationary within each segment and mitigates spectral leakage.
}

{
Accordingly, the minimum detectable coupling scales as $g_{\min}\left(m_a\right) \propto m_a^{1/2}$ in the $\tau_a$-limited regime, and becomes nearly mass independent when $T_{\text{seg}}<\tau_a$.
For wideband searches, we adopt the adaptive coherent segment above, $T_{\text{seg}}(\nu)=\min \left\{\varepsilon\tau_a(\nu), T_{\text{cap}}\right\}$, with $\varepsilon$ chosen to remain below the coherence time while respecting instrumental limits.
We then apply a frequency dependent look elsewhere penalty, estimating the number of trials as $N_{\text {trials }}(\nu) \sim \mathrm{BW} \times T_{\text{seg}}(\nu)$.
Throughout the analysis, power is summed optimally across $\Delta\nu_a$, avoiding the $\sqrt{N_{\text{bins}}}$ loss of single bin methods and preserving the $\sqrt{T_{\text{coh}}}$ scaling for each segment together with the $\sqrt{T_{\text{tot}}}$ gain from non-coherent stacking.
}

The spin precession dynamics induced by axion field have been investigated in different situations, including in strong electric field, and axion-photon interaction induced electron spin effective coupling \cite{Graham2013, Stadnik2014, Alexander2018, Zihang2021}.
Here we particularly focus on the spin polarization precession driven by the axion-induced effective magnetic field, tailored for silicon electron spin qubit as described in Ref.~\cite{Tan2025}, 
{
where the qubit array act as a sensitive magnetometers responding directly to the axion-induced effective magnetic field.
This approach avoids microwave circuits and tuning of the gate electric field required by electric spin resonance or AC Stark based schemes.
Moreover, the AC gate fields and charge defects can induce additional frequency shifts of the axion signal, complicating the natural triplet pattern.
}
Combining the qubit Hamiltonian and the interaction Hamiltonian together, we can obtain a longitudinal sideband frequency modulation with amplitude $\delta\omega=\gamma_e B_{\text{eff}}$, leading to:
\begin{equation}
    H(t) =\frac{1}{2} \bqty{ \omega_0 + \delta\omega \cos\theta(t) \cos(m_a t+\phi) } \sigma_z + \cdots \,.
\end{equation}
The measurable phase accumulation is proportional to the projection of the axion wind onto the qubit axis.
The factor $\cos\theta(t) = \hat{\bm{v}}_{\text{lab}}(t) \cdot \hat{\bm{q}}$ is modulated deterministically by Earth’s sidereal rotation and orbital motion \cite{Freese2013, Oka2017}.
FM sidebands will appear at $\omega=\omega_0 \pm n m_a$ with Bessel-law amplitudes $J_n(\beta)$ governed by the modulation index $\beta= \delta\omega / m_a$ (within a coherent segment where $\cos\theta(t)$ is quasi-static).
We employ a band-limited readout pipeline (resonator-assisted, down-conversion, and targeted filtering) that confines noise power while preserving the intrinsic sideband SNR.
The sensitivity scales as:
\begin{equation}
    \text{SNR} \sim \frac{B_{\text{eff}}}{\eta_B}\sqrt{t} \,,
\end{equation}
where $\eta_B$ in T/$\sqrt{\text{HZ}}$ is the sensitivity to a magnetic field, determined by the design of the qubit sensor, and $t$ is the coherent integration time.
This SNR provides a hardware-agnostic yardstick for comparing architectures (like NMR qubit, NV center qubit) and integration times.

The present work, to further reduce the impact of the device noise, investigate another unique feature of general dark matter searches: a parameter-free frequency spacing annual-modulation fingerprint in the baseband.
After selections of bands around $\omega_0 + m_a$ to form a baseband stream, this slowly varying geometric projection $\cos\theta(t)$ produces a sidereal day frequency at $\Omega_\star$ whose envelope is annually modulated.
Such a modulation will result in a robust sideband triplet at $\Omega=\{\Omega_\star, \Omega_\star\pm\Omega_\oplus\}$, with a fixed spacing $\Omega_\oplus$ and side-to-center amplitude ratio controlled by the annual depth.
This triplet is independent of $m_a$ (up to the heterodyning step) and thus complements the Bessel sidebands: together they provide two orthogonal signatures: mass-dependent high-frequency sidebands and mass-independent directional splitting, which can be used for joint detection and rejection.
Furthermore, because the daily and annual modulation phases $(\psi_\star,\psi_\oplus)$ are known from ephemerides, the coherent, heterodyned accumulation at $\{\Omega_\star,\Omega_\star \pm \Omega_\oplus\}$ will reveal the splitting even when the observation time is limited ($T_{\text{obs}}\ll 2\pi/\Omega_\oplus$), thus enabling sub-year campaigns.

In experimental design, we reuse the same front end as in the spectral engineering platform in Ref.~\cite{Tan2025}: a high-$Q$ resonator coupled to a CMOS-compatible spin qubit array \cite{Jakob2024}, Skipper cryo-CMOS down-conversion \cite{Quinn2023}, and narrowband digital filtering, so that the high-frequency analysis yielding FM sidebands (with amplitudes $\propto J_n(\beta)$) is preserved, while the baseband analysis tests the triplet at the known $\{\Omega_\star,\Omega_\star \pm \Omega_\oplus\}$.
The pair of signatures predict that common solar-day and line-power features do not naturally reproduce the fixed $\Omega_\oplus$ splitting, and the known amplitude ratios provide an internal consistency check.
The overall SNR accumulation follows the same scaling $\sim B_{\text{eff}}\sqrt{t}/\eta_B$, and improves with the state-of-art device fabrication including larger qubit number $N$, higher quality factor $Q$, and smaller $\eta_B$.

By considering a silicon electron spin qubit array as a narrowband, phase-coherent magnetometer, over one free precession window $\tau \ll 2\pi/\Omega_\star,\,2\pi/\Omega_\oplus$, the angle factor $\cos\theta(t)$ can be treated in the quasi-static limit.
The transverse quadrature follows the standard FM form:
\begin{equation}\label{Eq: localFM}
    \beta^{\text{loc}}=\frac{\delta\omega}{m_a} \cos\theta(t_0) \equiv \beta \cos\theta(t_0) \,.
\end{equation}
The spin polarization along the x-axis will precess as the following:
\begin{equation}
    \langle\sigma_x(t)\rangle = \cos \bqty{ \omega_0 t + \beta^{\text{ loc}} \sin(m_a t+\phi) + \varphi_0 } \,,
\end{equation}
producing Bessel-law sidebands at $\omega = \omega_0 + n m_a$ with amplitudes $\propto J_n(\beta^{\text{loc}})$.
To capture the slowly varying geometry, a fixed galactic wind direction $\hat{\bm{v}}_{\text{gal}}$ is rotated into the laboratory frame by Earth's motions:
\begin{equation}\label{Eq: rot_chain}
    \hat{\bm{v}}_{\text{lab}}(t) = R_\oplus (\Omega_\oplus t + \phi_\oplus) R_\star (\Omega_\star t+\phi_\star;\lambda) \hat{\bm{v}}_{\text{gal}} \,,
\end{equation}
where $R_\oplus$ and $R_\star$ are orbital rotation matrix and local sidereal rotation matrix respectively.
The angle factor $\cos\theta(t)$ can be expanded as:
\begin{equation}\label{Eq: beta_series}
\begin{aligned}
    &= c_0 + c_\star \cos  (\Omega_\star t - \psi_\star )  + c_\oplus\cos  (\Omega_\oplus t-\psi_\oplus ) + \\
    & + c_\times \cos  (\Omega_\star t-\psi_\star ) \cos  (\Omega_\oplus t-\psi_\oplus ) + \cdots \,.
\end{aligned}
\end{equation}
The constant coefficient $c_0$ is the time-averaged projection of $\hat{\bm{v}}_{\text{gal}}$ onto the qubit quantization axis, and $c_{\star}$, $c_{\oplus}$ and $c_{\times}$ are for daily modulation, annual modulation, and crossing modulation respectively.
All the expressions for $c_0,c_\star,c_\oplus,c_\times$ are determined by $\lambda, \hat{\bm{q}}, \hat{\bm{v}}_{\text{gal}}$.

Now we exam how these modulation is revealed in the qubit spectrum.
After band-selecting around $\omega_0 + n m_a$ to form a baseband stream $y(t)$, the daily component near $\Omega_\star$ is amplitude-modulated by the annual term.
The time-dependent amplitude of the daily-modulated signal $A_{\star}$ can be expressed as:
\begin{equation}\label{Eq: Astar}
    A_\star(t) = A_\star^{(0)} [1+\epsilon_\oplus\cos  (\Omega_\oplus t-\psi_\oplus ) ] \,,
\end{equation}
where $A_{\star}^{(0)} \propto c_{\star}$ denotes the baseline daily modulation amplitude (no annual effects).
This yields the baseband triplet $s_{\text{bb}}(t)$:
\begin{equation}\label{Eq: triplet_time}
\begin{aligned}
& A_{\star}^{(0)} \cos (\Omega_\star t-\psi_\star ) + \\
& \frac{1}{2} A_\star^{(0)} [\epsilon_\oplus \cos(\Omega_+ t - \psi_+ ) + \cos (\Omega_- t -\psi_-)] \,,
\end{aligned}
\end{equation}
where $\Omega_{\pm} = \Omega_\star \pm \Omega_\oplus$, $\psi_{\pm} = \psi_\star \pm \psi_\oplus$.
This is a clear indication of three baseband lines at three frequency with side-to-center amplitude ratio $\epsilon_\oplus$ (up to the finite-window line shape).
Furthermore, we notice that the spacing $\Omega_\oplus$ is set by Earth’s orbital frequency and is independent of instrumental parameters and $m_a$.

A finite dwell with taper $w(t)$ gives the following window response:
\begin{equation}\label{Eq: window}
    W(\Omega)=\int w(t) e^{- \text{i} \Omega t} \dd{t} \qquad  \delta f_W \lesssim \frac{\Omega_\oplus}{2\pi} \,,
\end{equation}
as the spectral resolvability condition for the triplet
($\delta f_W\simeq 1/T_{\rm obs}$ for a rectangular window).
Now, we form discrete, heterodyned statistics on time stamps $\{t_k\}$ from the band-limited stream $y_k = y(t_k)$:
\begin{equation}\label{Eq: stats}
    \mathcal{X}_\mu = \left|\sum_{k} w_k y_k e^{-\text{i} \Omega_\mu t_k} \right|^2 \,,
\end{equation}
where $\mu \in \{\Omega_{\star}, \Omega_+, \Omega_-\}$.
This $\mathcal{X}$  will coherently accumulate the split peaks even when $T_{\rm obs}\ll 2\pi/\Omega_\oplus$.
For stationary Gaussian noise with one-sided power spectral density (PSD),  $S(\Omega)$ can be evaluated locally.
Then the SNR can be estimated as following:
\begin{equation}\label{EQ: snr}
    \text{SNR}_\star \simeq  \frac{A_\star^{(0)}}{\sqrt{S(\Omega_\star)}}\sqrt{T_{\rm coh}} \quad 
    \text{SNR}_{\pm} \simeq \frac{\epsilon_\oplus}{2}\,{\rm SNR}_\star \,.
\end{equation}
Here, $T_{\text{coh}}$ is taken to be the minimal value among $T_2, \tau_c, T_{\text{obs}}$, where $\tau_c \simeq Q_a/m_a$ is the axion-field coherence time. And the spectral narrowness is characterized by the axion-field quality factor:
\begin{equation}
Q_a \equiv \frac{\omega_a}{\Delta \omega} \simeq \frac{2 c^2}{\left\langle v^2\right\rangle} \sim \mathcal{O}\left(10^6\right)
\end{equation}
Thus, the side peaks inherit a fixed fractional SNR set by $\epsilon_\oplus$, independent of the absolute sensitivity scale.

%========================================
%========================================
\section{Results and discussion}
\label{Sec: Results and discussion}

Before presenting quantitative results, we first specify the laboratory geometry and sensor orientations, which fix the deterministic sidereal and annual components of the projection $\cos\theta(t)=\hat{\bm{v}}_{\text{lab}}(t) \cdot \hat{\bm{q}}$.
Unless stated otherwise, all simulations adopt the DMGeometry instance summarized in Tab.~\ref{Tab: Geom}, corresponding to a Beijing site and a zenith-pointing device with zero turntable rotation and zero initial local sidereal time (LST).
These choices reproduce the code snippet used to generate our baselines and set the reference configuration for the triplet analysis discussed below.

%==========
\begin{table}[htbp!]
\begin{ruledtabular}
\begin{tabular}{lll}
Parameter  & Description & Value \\
\hline
$\lambda$   & geographic latitude   & $39.9042^{\circ}$ \\
$\varphi$   & geographic longitude  & $116.4074^{\circ}$  \\
$\alpha_{\text{w}}$    & wind right ascension  & $270^{\circ}$ $(3\pi/2)$ \\
$\delta_{\text{w}}$    & wind declination  & $ 30^{\circ}$ $(\pi/6)$ \\
$\text{elev}$   & device elevation  & $90^{\circ}$ $(\pi/2)$ \\
$\text{azim}_0$ & device azimuth    & $0^{\circ}$ $(0)$ \\
$\dot\varphi_{\text{turn}}$    & turntable angular rate    & $0 \text{rad s}^{-1}$ \\
$\text{LST}_0$    & initial local sidereal phase  & $0$ \\
\end{tabular}
\end{ruledtabular}
\caption{Geometry parameters used in the simulations (DMGeometry instance).
Angles are shown in degrees; radian forms are given in parentheses.
For Beijing, we use $\lambda\simeq 39.9042^{\circ}$ (N) and $\varphi\simeq 116.4074^{\circ}$ (E).
\label{Tab: Geom}
}
\end{table}
%==========

%==============================
%==============================
\subsection{Fingerprints of the axion annual modulation}

\begin{figure}[htb]
\centering
\includegraphics[width=\columnwidth]{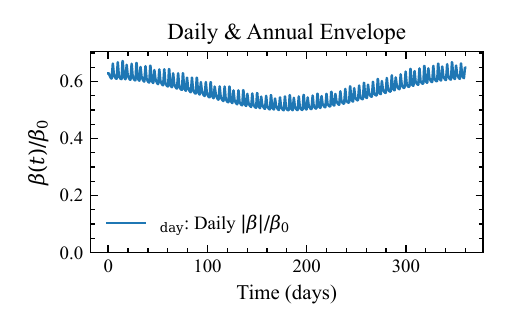}
\caption{
\textbf{Daily and annual envelope of the geometric modulation}.
Shown is the instantaneous daily series of $|\beta(t)|/\beta_0 = (v_{\rm lab}(t)/v_0)|\cos\theta(t)|$ over a full year (site latitude and sensor orientation as indicated in Table~\ref{Tab: Geom}).
The slow variation of the envelope follows the annual modulation at $\Omega_\oplus/2\pi \simeq 3.17\times10^{-8}\,\mathrm{Hz}$, while the rapid ripples are set by Earth’s \emph{sidereal} rotation at $\Omega_\star/2\pi \simeq 1.1606\times10^{-5}\,\mathrm{Hz}$.
Here $\beta_0$ fixes the overall scale (set by $g_{ae}\sqrt{2\rho_a}$ and a reference speed $v_0$).
}
\label{Fig: envelope}
\end{figure}

In Fig.~\ref{Fig: envelope}, we plot the purely geometric component of the signal.
At a fixed calendar day $t$, we define the daily projection $\beta(t')/\beta_0\equiv\cos\theta(t,t')$, where $t'$ runs over one sidereal day.
Using Eq.~\eqref{Eq: beta_series}, the projection can be written as:
\begin{align}
    & \cos\theta(t,t') = \mu_{\text{d}}(t) + K(t) \cos(\Omega_\star t'-\psi_\star) \\
    & \mu_{\text{d}}(t) = c_0+c_\oplus\cos(\Omega_\oplus t-\psi_\oplus) \\
    & K(t) = c_\star+c_\times\cos(\Omega_\oplus t-\psi_\oplus) \,.
\end{align}
The {blue line} in Fig.~\ref{Fig: envelope} is simply the envelope:
\begin{equation}
    \cos\theta_{\max/\min}(t)=\mu_{\rm d}(t)\pm|K(t)| \,,
\end{equation}
which is the daily mean $\mu_{\text{d}}(t)$ slowly drifts over the year, while the daily excursion amplitude $|K(t)|$ is annually modulated.
The frequency of the slow drift is fixed by celestial parameter $\Omega_\oplus$, and the overall scale $\beta_0$ depends on $g_{ae}\sqrt{2\rho_a}$ and the time-averaged projection, but not on the axion mass.

%==========
%==========

The envelope in Fig.~\ref{Fig: envelope} connects directly to the baseband triplet of Eq.~\eqref{Eq: triplet_time}. The daily waveform is an amplitude-modulated signal:
\begin{equation}
    \frac{\beta(t)}{\beta_0}= [1+\epsilon_\oplus\cos(\Omega_\oplus t+\psi)][\mu_{\rm d}+K\cos(\Omega_\star t+\phi)] \,,
\end{equation}
where the two terms are for annual and diurnal term respectively.
Expanding to leading order in the small annual depth $\epsilon_\oplus$ makes the annual $\times$ diurnal product equivalent (in frequency space) to the linear superposition of a carrier at $\Omega_\star$ and two yearly sidebands at $\Omega_\star \pm \Omega_\oplus$, with sideband-to-carrier amplitude ratio $\epsilon_\oplus/2$.
In the time domain this appears exactly as a fast diurnal oscillation whose amplitude slowly breathes over the year.

%==========
\begin{figure}[htb]
\centering
\includegraphics[width=\columnwidth]{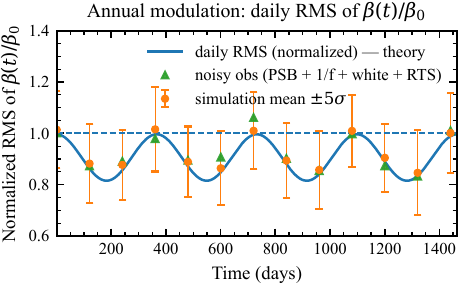}
\caption{
\textbf{Daily RMS and its annual modulation (theory vs. noisy observations).}
For each sidereal day we compute $\mathrm{RMS}\{\beta(t)/\beta_0\}$ and normalize it by the yearly mean (vertical axis).
The solid curve shows the geometry-only prediction; {Orange circles denote simulated noisy observations generated with Pauli spin blockade type readout and simulated $1/f$, white, and random-telegraph noise.  
Green triangles indicate the Monte Carlo mean $\pm5\sigma$ from the same noise ensemble, and the blue solid line shows the geometry-only theoretical prediction.}
The slow envelope oscillates at the annual frequency $\Omega_\oplus/2\pi \simeq 3.17\times10^{-8}\,\mathrm{Hz}$ and fixes the sideband-to-carrier amplitude ratio of the $\Omega_\star\pm\Omega_\oplus$ triplet [Eq.~(\eqref{Eq: triplet_time})], namely $\epsilon_\oplus/2$.
Note that the normalization uses a reference $\beta_0$ (set by $v_0$), so values may exceed unity when $v_{\rm lab}>v_0$.
}
\label{Fig: dailyRMS}
\end{figure}
%==========

Figure~\ref{Fig: dailyRMS} compares the geometry-only prediction with simulated noisy observations (the noise configurations can be found in Ref.\cite{Tan2025}). Error bars represent the expected $5\sigma$ scatter of experimental data rather than the statistical uncertainty of the simulation itself. For each sidereal day we form $R_{\mathrm{d}}(t)\equiv \mathrm{RMS}\{\beta(t)/\beta_0\}$ and normalize it by the yearly mean. And the sidereal tone at $\Omega_\star$ has an amplitude proportional to $|K(t)|$, expanding $K(t)$ to leading order in the small annual term gives a near-sinusoidal modulation at $\Omega_\oplus$.
Equivalently, the daily RMS will read:
\begin{equation}
    R_{\text{d}}(t)=\sqrt{ \frac{1}{T_{\text{sid}}} \int_t^{t+T_{\text{sid}}} \cos^2\theta(t',t) \dd{t'} } \,,
\end{equation}
which evaluates to $\sqrt{\mu_{\text{d}}^2(t) + K^2(t)/2}$ will inherit the same annual envelope.

After normalization, this envelope fixes the side-to-center amplitude ratio of the triplet and is independent of $m_a$, providing a mass–independent, geometry-fixed discriminator. The markers are generated with a Pauli spin blockade like (non-Gaussian) readout model and injected $1/f$, white, and random-telegraph noise (error bars show the Monte-Carlo mean $\pm k\sigma$), while the solid curve is the noise-free geometric expectation.
As anticipated from the expansion in Eq.~\eqref{Eq: beta_series}, $R_{\mathrm{d}}(t)$ retains a near-sinusoidal annual envelope at $\Omega_\oplus$, with small geometry-induced deviations from the mixed term.
This mass-independent envelope fixes the relative amplitudes of the baseband triplet at $\{\Omega_\star,\ \Omega_\star \pm \Omega_\oplus\}$ in Eq.~\eqref{Eq: triplet_time} (sideband-to-carrier ratio $\epsilon_\oplus/2$), and the agreement between the noisy points and the theory curve demonstrates robustness to realistic readout noise.

%==========
\begin{figure}[htb]
\centering
\includegraphics[width=\columnwidth]{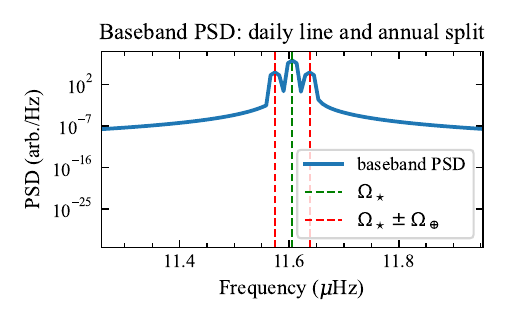}
\caption{
\textbf{Baseband power spectral density and the annual-splitting triplet}.
PSD around the sidereal line at $\Omega_\star$ exhibits a triplet at 
$\{\Omega_\star,\ \Omega_\star \pm \Omega_\oplus\}$ produced by annual amplitude modulation of the daily component. 
Green dashed line marks $\Omega_\star$; red dashed lines mark $\Omega_\star \pm \Omega_\oplus$. 
{The simulated precision $\delta f_W=7.93\times10^{-9}$Hz is the finite simulation resolution and windowing of the estimator}; ideally the separation equals $\Omega_\oplus/2\pi\simeq3.17\times10^{-8}$ Hz. 
The vertical scale is in arbitrary units per Hz; see text for the estimator (Welch/taper) and dwell time.
}
\label{fig:psd}
\end{figure}
%==========
Figure~\ref{fig:psd} zooms into the baseband around the sidereal line.
As predicted by Eq.~\eqref{Eq: triplet_time}, the PSD exhibits a triplet at 
$\{\Omega_\star,\ \Omega_\star \pm \Omega_\oplus\}$: the central daily line (green dashed) 
and two annually split companions (red dashed). 
The frequency spacing is fixed by Earth's orbital motion.
In the ideal limit, we have used $\Omega_\oplus/2\pi \simeq 3.17\times 10^{-8}$ Hz, independent of the axion mass or instrument details.

The blue curve is the windowed periodogram (Welch/taper); each line is the true spectral line convolved with the window response $|W(\Omega)|^2$.
For a segment length $T_{\text{seg}}$, the effective resolution bandwidth is $\delta f_W \simeq 1/T_{\text{seg}}$ (Hann: $\sim 1.44/T_{\text{seg}}$), so resolvability requires $\delta f_W \lesssim \Omega_\oplus/2\pi$.
{If $\delta f_W\gtrsim \Omega_\oplus/2\pi$, the three lines partially blend, which is undesired.}

The relative heights of the split peaks carry purely geometric information: to leading order the side peaks are $(\epsilon_\oplus/2)$ of the central daily line after accounting for the common window.
This pattern is mass independent and serves as a directional discriminator: solar-day features near $\Omega_\odot/2\pi \simeq 1.16 \times 10^{-5}$ Hz (solar-day frequency) and power-line harmonics do not reproduce a symmetric split at the fixed spacing $\Omega_\oplus/2\pi \simeq 3.17 \times 10^{-8}$ Hz.
In the regime where the peaks are not fully resolved, we instead rely on the ephemeris-guided narrowband statistics of Eq.~\eqref{Eq: stats} (matched filtering/heterodyning at $\Omega_\star$ and $\Omega_\star \pm \Omega_\oplus$), which coherently accumulate the triplet without requiring a full-year dwell.

%==============================
%==============================
\subsection{Estimated detecting region}

\begin{figure}[htb]
  \centering
  \includegraphics[width=\columnwidth]{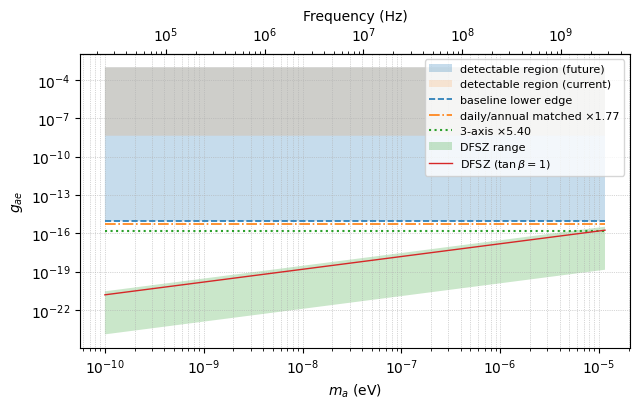}
  \caption
  {
    \textbf{Axion–electron coupling sensitivity including geometric enhancements}.
    Shaded areas indicate the detectable regions for the current and future configurations, while the band shows the DFSZ model range with the $\tan\beta=1$ benchmark curve.
    Three lower-edge curves are overlaid: the baseline limit (single axis, uniform weighting), the limit with matched weighting to daily/annual modulation, and the limit with matched weighting plus three-axis readout.
  }
  \label{Fig: gaee-sensitivity-enhanced}
\end{figure}

To quantify the geometric enhancements, we consider a site at latitude $\phi=39.9^\circ$ (Beijing) with the DM-wind declination $\delta=30^\circ$.
The time-averaged projection of the wind direction onto a zenith-pointing qubit axis is $p_0=\sin\phi\,\sin\delta=0.321$, while the mean-square projection evaluates to $\langle P^2\rangle = p_0^2 + (\cos\phi\cos\delta)^2/2 = 0.324$.
These quantities give us the geometric daily and annual matched-weighting gains respectively: $G_{\text{ daily}}=\sqrt{\langle P^2\rangle}/|p_0| \simeq 1.77$, as well as the three-axis readout gain (relative to the single-axis matched case) $G_{{\text{3axis}}} = (\sqrt{\langle P^2\rangle})^{-1} \simeq 1.76$
Combining these factors with the standard quantum limit scaling for three identical sensors ($N_{\text{total}}=3N$, resource factor $\sqrt{N}=\sqrt{3}$), the overall enhancement relative to a baseline single-axis, uniformly weighted array is $G_{\text{total}} = \sqrt{N} G_{\text{daily}} G_{\text{3axis}} \simeq 5.40$. {Fig.~\ref{Fig: gaee-sensitivity-enhanced} shown the $5\sigma$ axion searching sensitivity with the geometric enhancements, the SNR estimation and noise modeling for the searching sensitivity can be found in Ref.~\cite{Tan2025}, and the detailed parameters for the electron spin qubits and axion wind follow the Table~\ref{tab:parameters}.}

\begin{table}[htb]
\caption{Parameters used in the axion-search sensitivity estimation.}
\centering
\begin{tabular}{ccc}
\hline\hline
\textbf{Symbol} & \textbf{Description} & \textbf{Value} \\
\hline
$\rho_a$ & Local dark matter density & $0.4~\mathrm{GeV/cm^3}$ \\
$v_a$ & Axion velocity & $1\times10^{-3}\,c$ \\
$\gamma$ & Spin gyromagnetic ratio & $28\times10^{9}~\mathrm{Hz/T}$ \\
$T_1$ & Longitudinal relaxation time & $1~\mathrm{ms}$ \\
$T_2$ & Transverse dephasing time & $100~\mathrm{\mu s}$ \\
$B_0$ & Static magnetic field & $0.5~\mathrm{T}$ \\
$N_{\mathrm{current}}$ & Number of spins (current) & $10$ \\
$N_{\mathrm{future}}$ & Number of spins (future) & $10^{6}$ \\
$Q$ & Resonator quality factor & $10^{4}$--$10^{6}$ \\
\hline\hline
\end{tabular}
\label{tab:parameters}
\end{table}

%==========
%==========

However, {Fig.~\ref{Fig: gaee-sensitivity-enhanced}} above estimate ignores the axion mass $m_a$ and the coherence time $\tau_a$ indicating that the detection sensitivity is mass-independent and neglects spectral broadening. In addition to geometric considerations, the astrophysical linewidth of the axion field under the SHM must be incorporated as mentioned in Eqs.~\eqref{Eq: v2-closed}-\eqref{Eq: tau_a}. We have calculated that the fractional linewidth is $\Delta\nu_a/\nu_a \simeq 3.9 \times 10^{-7}$ and the corresponding coherence time $\tau_a^{-1} = \pi \Delta \nu \propto m_a$.
%==========
\begin{figure}[htb]
\centering
\includegraphics[width=\columnwidth]{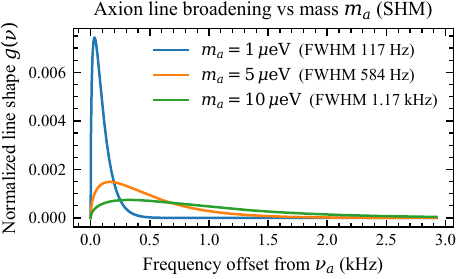}
\caption{
\textbf{Axion line broadening as a function of mass $m_a$ (SHM).}
Shown are the normalized Standard Halo Model line shapes $g(\nu)$
for $m_a=\{1,5,10\}\,\mu\mathrm{eV}$ versus the frequency offset from $\nu_a$.
The absolute width (FWHM) scales linearly with $m_a$ (e.g.\ $\approx117~\mathrm{Hz}$ at $1~\mu\mathrm{eV}$),
whereas the relative width is mass–independent:
$\Delta\nu/\nu_a \simeq v_0^2/(2c^2)\approx2.7\times10^{-7}$,
corresponding to $Q_{\rm FWHM}\!\sim\!2.1\times10^6$.
The SHM shape has support only for $\nu\ge\nu_a$; detector windows and coherence-time effects
(\,$\tau_c=1/(\pi\Delta\nu)$\,) enter multiplicatively in sensitivity estimates.
}
\label{Fig: axion_broadening}
\end{figure}
%==========

\begin{figure}[htb]
  \centering
  \includegraphics[width=\columnwidth]{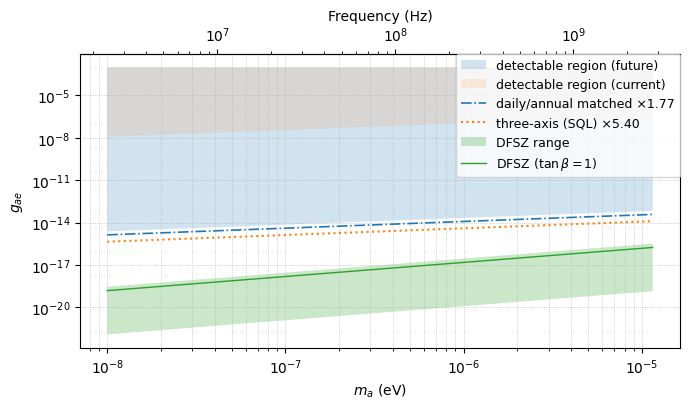}
  \caption{
  \textbf{Sensitivity to the axion–electron coupling including realistic SHM coherence and an adaptive coherent segment}.
  Shaded areas show detectable regions for current and future configurations, with the green band indicating the DFSZ range.
  Lower-edge curves illustrate successive improvements: baseline, matched daily/annual weighting ($\times1.77$), and three-axis readout with standard quantum limit scaling ($\times\sqrt{3}$).
  The combined geometric and resource gain reaches $\simeq5.40$ for $\phi=39.9^\circ$, $\delta=30^\circ$.
  A global false-positive rate of $\alpha=1\%$ is enforced with a look–elsewhere corrected threshold.
  }
  \label{Fig: gaee-sens-shm-3axis}
\end{figure}

Fig.~\ref{Fig: axion_broadening} illustrates that while the absolute FWHM grows $\propto m_a$, the fractional width, and thus the axion quality factor is essentially fixed by the halo kinematics, $\Delta\nu/\nu_a \simeq v_0^2/(2c^2)$.
So in the sensitivity calculation below, we adopt an adaptive coherent segment length $T_{\text{seg}}(\nu)=\min\{\epsilon\tau_a(\nu), T_{\rm cap}\}$, where $\epsilon$ is an experimental safety factor and $T_{\text{cap}}$ is the maximum capture time, and apply a frequency-dependent look–elsewhere correction $z_{\alpha/N_{\text{trials}}(\nu)}$ with $N_{\text{trials}}(\nu) \sim \text{BW} T_{\text{seg}}(\nu)$.
Here $\alpha$ denotes the global false-positive rate (significance level) for the entire search, typically fixed at $\alpha=1\%$ unless otherwise stated. Within each axion linewidth $\Delta\nu_a$, the signal power is summed optimally, thereby avoiding the $\sqrt{N_{\text{bins}}}$ degradation associated with a single-bin pickoff.
The resulting sensitivity curves therefore exhibit the correct $m_a$ dependence: nearly flat when $T_{\text{seg}}<\tau_a$, and scaling as $g_{\min}\propto m_a^{1/2}$ in the $\tau_a$-limited regime.
Fig.~\ref{Fig: gaee-sens-shm-3axis} summarizes these results with the consideration of Axion line broadening.  
The baseline limit already reflects the SHM coherence and adaptive segmentation, while the overlaid curves demonstrate the gains from daily/annual matched weighting ($\times G_{\text{daily}}$) and three-axis readout ($\times G_{\text{3axis}}$), with an additional SQL factor of $\sqrt{3}$ from parallel sensors.  
Altogether, the combined enhancement relative to the baseline reaches $G_{\text{total}}\simeq 5.40$ for the chosen geometry.

%==========
{
The current region corresponds to a 10-qubit silicon spin–qubit sensor (single-axis readout), with coupled High-Q resonator $Q \approx 10^{4}$ for the electron spin qubit, and magnetic sensitivity $\eta_B \approx 10^{-15}$ T/Hz \cite{Szechenyi2017, Holman2020, Gonzalez2021}. 
The future region represents a three-axis, $10^{6}$-qubit array operated near the standard quantum limit, with $Q \approx 10^{6}$, per-qubit $\eta_B \approx 10^{-16}\,\mathrm{T}/\sqrt{\mathrm{Hz}}$. 
Both cases use the same SHM linewidth, the adaptive coherent segmentation $T_{\mathrm{seg}}(\nu)=\min\{\varepsilon\,\tau_a(\nu),T_{\mathrm{cap}}\}$, optimal power-summing within the axion linewidth, and ephemeris-guided baseband triplet statistics.
}

{
Both regimes share the same signal-processing chain—adaptive coherent segmentation $T_{\mathrm{seg}}(\nu)$, optimal power-summing within the axion linewidth, and the baseband triplet statistics derived from the sidereal and annual modulation fingerprints.
The improvement from the current to the future case therefore stems entirely from hardware parameters (qubit number, readout geometry, and resonator $Q$) rather than from any modification of the data-analysis method.  
These results indicate that the proposed approach can scale naturally with advances in semiconductor quantum hardware, providing a realistic path toward probing the astrophysically motivated range of axion–electron couplings.}
%==========

%========================================
%========================================
\section{Conclusions}
\label{Sec: Conclusions}

We have identified a parameter–free, geometry–fixed annual–modulation fingerprint for axion–wind searches with silicon spin qubits: a baseband triplet at $\Omega_\star$ and $\Omega_\star\pm\Omega_\oplus$.
Its spacing is set solely by celestial mechanics and the side–to–center amplitude ratios are fixed by the daily–RMS envelope of $\cos\theta(t)$ (site latitude and instrument pointing).
This makes the morphology of the triplet independent of the axion mass and therefore a robust directional diagnostic complementary to the carrier–scale FM sidebands at $\omega_0 \pm n m_a$ governed by $\beta=\delta\omega/m_a$.

On the modeling side, we incorporated the astrophysical linewidth of the axion field under the SHM, which yields $\Delta\nu_a/\nu_a \simeq 3.9 \times 10^{-7}$ and a coherence time $\tau_a \propto 1/m_a$.
This introduces the correct $m_a^{1/2}$ scaling of the sensitivity in the $\tau_a$–limited regime and motivates the use of an adaptive coherent segment $T_{\text{seg}}(\nu)=\min{\epsilon, \tau_a(\nu), T_{\rm cap}}$.
Together with optimal power summation across the linewidth and frequency–dependent trial correction, this produces physically consistent sensitivity curves.
On top of this baseline, geometric matched weighting ($\times1.77$), three–axis readout ($\times1.76$), and standard quantum limit resource scaling ($\times\sqrt{3}$) combine to an overall enhancement of $\sim 5.4$ for our site geometry. The resulting pipeline coherently recovers the $\Omega_\oplus$ splitting without requiring year-long dwell, improves rejection of stationary/instrumental backgrounds, and enables joint inference of coupling strength and sensor orientation.

Looking ahead, the same framework can accommodate non-SHM velocity structures (streams, debris flow, disk) \cite{Christy2024, Herrera2025}, and the triplet statistic together with adaptive coherence can be integrated to other spin-based sensors like germanium hole spin qubits and multi-site arrays \cite{Watzinger2018, Terrazos2021, Hendrickx2020}.
This opens the way to falsifiable, cross-validated axion–wind searches with semiconductor quantum dot spin qubits on sub-year timescales.

\newpage
\bibliography{Bibitem}
\end{document}